# Simultaneous, inherently temperature and strain insensitive bio-sensors based on dual-resonance long-period gratings

Saurabh Mani Tripathi*[a], Deep Shikha Verma[a], Wojtek J. Bock[b], Predrag Mikulic[b]
[a]Department of Physics, Indian Institute of Technology Kanpur, U.P., India 208016
[b]Photonics Research Center, University of Quebec at Outaouais, Gatineau, QC, Canada

## ABSTRACT

Addressing temperature and strain induced cross-talks simultaneously, we propose an inherently strain and temperature insensitive fiber-optic bio-sensor. The insensitivity has been achieved by properly adjusting the dopants and their concentrations, and by optimizing the grating period and the strength of concatenated dual-resonance long-period-gratings.

## 1. INTRODUCTION

Cross talk with respect to the unwanted perturbations are the biggest issues to be addressed for any reliable sensor system. Two primary causes of error related to the fiber-optic sensor systems are temperature fluctuations and variations in the axial strain of the fiber-optic transducers. Often, a simultaneous measurement of these parameters is necessary to filter out the response of the target parameter to be sensed[1-3]. Recently, some attempts have also been made to make the sensor insensitive with respect to some of these parameters[4-6]. However, most of these methods are characteristically extrinsic in nature, which often involves thermal insulation packaging and/or specific coating of the sensor making the evanescent wave associated with various modes well buried inside the coating. Such an approach removes any possibility of bio-sensing. Recently we have proposed and demonstrated an inherently temperature insensitive bio-sensor utilizing two concatenated long-period gratings[6]. The insensitivity had been achieved by judiciously exciting cladding modes by means of partially etched dual-resonance long-period gratings (DRLPGs) and optimizing an inter-grating space where the excited cladding mode supports single resonance. Although the sensor showed excellent temperature insensitivity with respect to the changing temperatures, cross-sensitivity with respect to the axial strain was still present. Further, the overall sensor length was also high which is not desirable for bio-sensing applications.

In this paper, we propose a new bio-sensor that is simultaneously, inherently insensitive to the temperature as well as to axial strains applied to the sensor, and has much smaller grating length as compared to earlier temperature insensitive sensor.

## 2. SENSOR PARAMETER OPTIMIZATION

Changes in both the temperature as well as in the axial strain, has a two fold effect on the modal properties of an optical fiber, it changes (*i*) the diameter of the optical fiber, and (*ii*) refractive index of the core as well as cladding regions. Both of these effects, in turn, change the effective indices of the core and cladding modes involved in the grating assisted coupling. The expansion/contraction of the optical fiber further affects the grating period, changing the resonance wavelength considerably. One way to inherently nullify the effects of temperature is to excite the dual-resonance and single-resonance cladding modes and make them interfere with the core mode[6]. However, using this method only one side of the transmission spectrum (typically on the higher wavelength side of the turn-around wavelength) can be made temperature insensitive. In our present work we follow a new approach of properly adjusting the dopants in the optical fiber core-region to achieve an overall temperature insensitivity. Typically, to increase the refractive index of the core region, a fraction of $GeO_2$ is doped in an otherwise $SiO_2$ core. Both, the $GeO_2$ as well as $SiO_2$ have positive thermo-optic coefficients (dn/dt ~$1.06 \times 10^{-5}$/°C for fuzed $SiO_2$, and $1.24 \times 10^{-5}$/°C for 15 mole% $GeO_2$ doped $SiO_2$). Further, in order to reduce the softening temperature of the fiber as well as to increase its photo-sensitivity at UV radiation, the optical fiber core region is also co-doped with $B_2O_3$ which generally reduces the core regions refractive index[7]. Apart from reducing the core regions refractive index, owing to its negative thermo-optic coefficient (dn/dt ~ -3.5 × $10^{-5}$/°C ) $B_2O_3$ also reduces the overall thermo-optic coefficient of the fiber core region. It appears that by properly adjusting the doping concentrations of these two dopants in the fiber core region, an overall temperature insensitive DRLPG can be realized.

*smt@iitk.ac.in; phone +91-512-259-6871; fax 91 512 259-0914

In order to numerically study this, we first simulated transmission spectrum of an LPG operating in its highly sensitive dual-resonance regime. In our simulations we first considered an optical fiber with its core region made of 4.1 mole% $GeO_2$ doped $SiO_2$, and its diameter 8.2 μm. The cladding region is considered to be made of fused $SiO_2$, and its diameter 125 μm. These opto-geometric parameters are chosen in such a way that the modal characteristics of the fiber are close to that of Corning SMF-28™ optical fiber[6,8]. The $LP_{011}$ cladding mode of such fiber shows its turn-around wavelength around 1625 nm, which matches very well our earlier experiments. The field distribution of this mode is shown in Fig.1(a), and the corresponding transmission spectrum, with the LPG subjected to different ambient refractive indices (ARI), temperature and strain, for the power coupling between $LP_{01}$ core mode to $LP_{011}$ cladding mode is shown in Fig.1(b). Evidently, the LPG is extremely sensitive to the changes in the ARI, temperature and axial strain. Defining sensitivity as $S = \frac{\Delta|\lambda_{R1} - \lambda_{R2}|}{\Delta\chi}$ where $\lambda_{Rm}$ is the resonance wavelength of the $m^{th}$ resonance and $\chi$ is the measurand quantity, the ARI, temperature and axial strain sensitivities are 2430 nm/RIU, 1.78 nm/°C and 4 pm/με, respectively, which can be further increased by bringing the two resonance wavelengths closer to each other by means of increased grating period and/or grating strength. Having estimated the sensitivity of the sensor with respect to different external parameters we slightly changed the composition of the core region by introducing small concentration of $B_2O_3$ in the binary combination of $GeO_2$ doped $SiO_2$ core. The refractive index of the ternary glass (core region) is obtained using the Maxwell–Garnet (MG) effective medium approach[9]:

$$\sum_m \left(\frac{\varepsilon_m - \varepsilon_t}{\varepsilon_m + 2\varepsilon_t}\right) v_m = 0 \qquad (1)$$

where, $\varepsilon_m (= n_m^2)$ and $v_m$ are, respectively the dielectric constant and the volume fraction of the of $m^{th}$ dopant. The MG approach gives better agreement with the experimental values (within 1.9%) as compared to the Lorentz–Lorentz equation[10]. The transmission spectrum of the LPG written in the ternary core with a composition of 1.5 mole% $B_2O_3$, 4.1 mole% $GeO_2$ in 94.4 mole% $SiO_2$ at three different temperature is shown in Fig.2(a), showing opposite nature of spectral shifts as compared to Fig. 1(b). This clearly indicates that for a suitable concentration of $B_2O_3$ in the optical fiber core region, the LPG should be insensitive to variations in temperature. This has been shown in Fig. 2(b) where we have plotted the transmission spectrum using a core composition of 0.75 mole% $B_2O_3$, 4.1 mole% $GeO_2$ in 95.15 mole% $SiO_2$ at different temperature; complete temperature insensitivity throughout the 1.44-1.77 μm wavelength range is obtained for this core composition. The grating period, strength and length are 174.2 μm, $0.44\times10^{-4}$ and 3.2 cm, respectively.

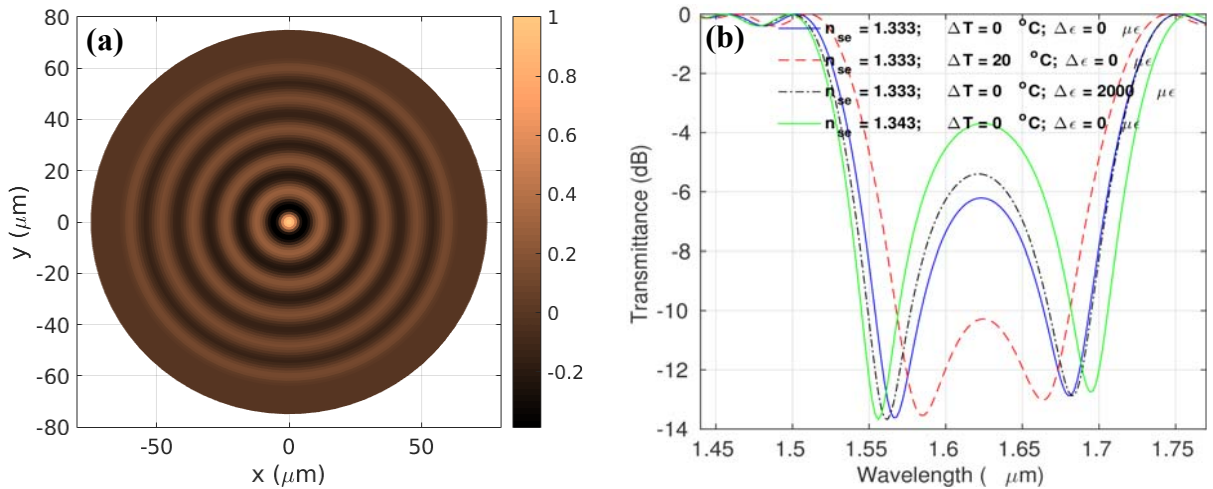

Figure 1. (a) Field distribution of $LP_{011}$ cladding mode, the turn-around wavelength corresponding to $LP_{01}$ – $LP_{011}$ mode power coupling is at 1.625 μm. (b) transmission spectrum with varying ARI, temperature and axial strain.

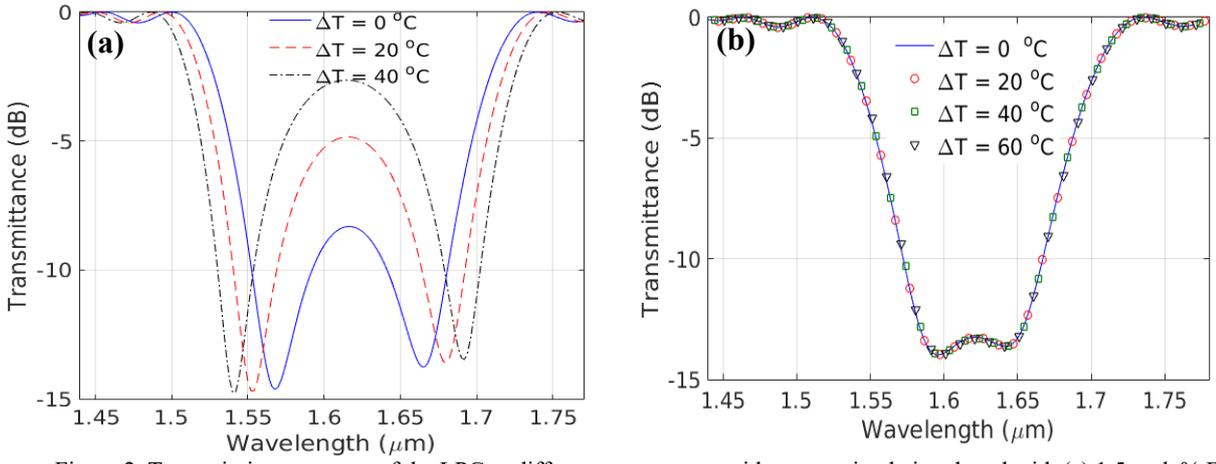

Figure 2. Transmission spectrum of the LPG at different temperature with core region being doped with (a) 1.5 mole% $B_2O_3$, and (b) 0.75 mole% $B_2O_3$. The host material is 4.1 mole% $GeO_2$ in $SiO_2$.

*smt@iitk.ac.in; phone +91-512-259-6871; fax 91 512 259-0914

## 2.1 Simultaneous strain and temperature insensitivity

Having obtained an overall temperature insensitivity, we now focus on the axial strain. Keeping in view that the two resonance wavelengths of the DRLPG move opposite to each-other, one of the resonances can be made insensitive to a given external perturbation by making it to coincide with the other resonance by means of two DRLPGs. Here it is important to mention that in order to avoid modal interference between various cladding modes and/or between the core and cladding modes, there should not be any inter-grating space between the two LPGs. Schematically this has been shown in Fig.3, in which two DRLPGs are concatenated without any inter-grating space between them. The grating strengths (GS) of these LPGs is optimized in such a way that the higher resonance wavelength of first DRLPG coincides with the lower resonance wavelength of the second DRLPG. The overall transmission spectrum of this sensor is plotted in Fig.4(a), showing a complete temperature insensitivity throughout the entire spectral range, as well as a strain insensitive higher resonance wavelength ($\lambda_{R2}$). The grating period of both the DRLPGs are taken as 174.2 μm and the GS of the individual

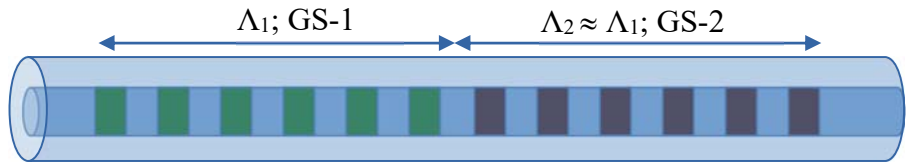

Figure 3. Schematic diagram of the sensor consisting two concatenated DRLPGs with different grating strength

LPGs are taken as $0.44\times10^{-4}$ and $1.76\times10^{-4}$ respectively, the individual grating lengths are 2 cm. In Fig.4(b) we have plotted the variation of resonance wavelength separation with varying ambient refractive indices. The refractive index sensitivity $\left(S = \frac{\Delta|\lambda_{R1} - \lambda_{R2}|}{\Delta n_{se}}\right)$ being 4607 nm/RIU. We would like to mention that the ARI sensitivity is increased due to (*i*) reduced separation between the two resonance wavelength (sensitivity is effectively infinite precisely at the turn-around wavelength due to vanishing slope of the Λ vs λ curve), and (*ii*) slight increase in the turn-around wavelength (owing to the increased GS of the second LPG).

*smt@iitk.ac.in; phone +91-512-259-6871; fax 91 512 259-0914

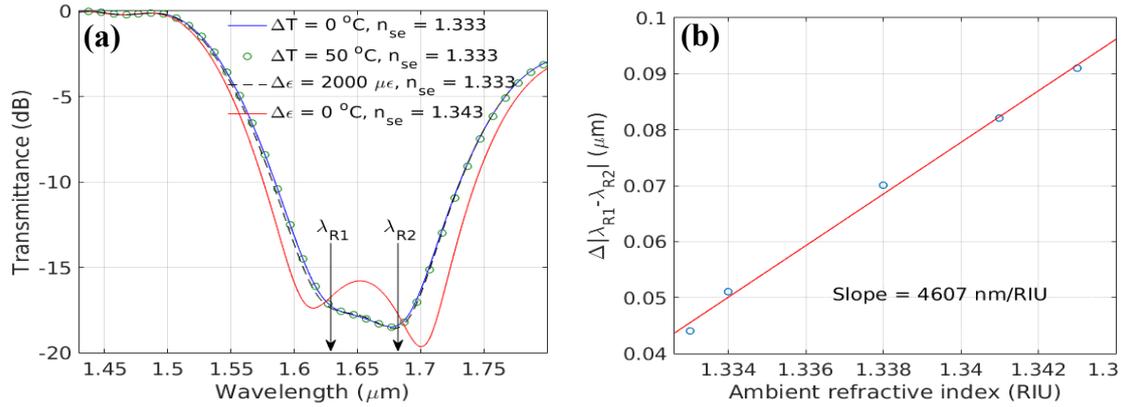

Figure 4. Transmission spectrum of the sensor at two different ARI, temperature and axial strains. (b) variation of the resonance wavelength separation with changing ambient refractive indices; the ARI sensitivity is 4607 nm/RIU.

## 3. CONCLUSIONS

In conclusion, in this paper we have presented a simultaneous, inherently temperature and axial strain insensitive bio-sensor. Temperature insensitivity is obtained by optimizing the core dopants and their concentration, whereas a simultaneous strain insensitivity has been obtained by suitable selecting the grating period and strength of the dual-resonance LPGs so that the higher resonance wavelength of one DRLPG falls on the lower resonance wavelength of the other DRLPG. This can be achieved by properly adjusting the exposure time of the individual DRLPGs. The resultant sensor shows an ARI sensitivity of 4607 nm/RIU, which can be used to detect changes as small as $2.2 \times 10^{-7}$ in ambient refractive indices using a detection system with spectral resolution of 1 pm. The sensor is currently under fabrication at our lab.

## 4. ACKNOWLEDGEMENTS

The authors gratefully acknowledge supports from the NSERC and CRC programs of Canada. SMT and DSV also acknowledge support from DST, Govt. of India through an INSPIRE project (grant no. IFA13-PH-69).

*smt@iitk.ac.in; phone +91-512-259-6871; fax 91 512 259-0914